\documentclass[aps,prl,twocolumn,showpacs,longbibliography,floatfix]{revtex4-1}
\usepackage{graphicx}
\usepackage{amsmath}
\usepackage{amssymb}
\usepackage{bbold}
\usepackage{color}
\usepackage[normalem]{ulem}
\usepackage{epsfig}
\usepackage{float}

\begin{document}
\title{A generalized model  of island biodiversity}
\author{David A. Kessler and Nadav M. Shnerb}
\affiliation{Department of Physics, Bar-Ilan University, Ramat-Gan
IL52900, Israel}
\begin{abstract}
The dynamics of a local community of competing species with weak
immigration from a static regional pool is studied. Implementing the generalized competitive Lotka-Volterra model with 
demographic noise, a rich dynamics structure with four qualitatively distinct phases is unfolded. When the overall interspecies competition is weak, the island species are a sample of the mainland species. For higher values of the competition parameter the system still admit an equilibrium community, but now some of the mainland species are absent on the island.  Further increase in competition leads to an intermittent "chaotic" phase, where the dynamics is controlled by invadable combinations of species and the turnover rate is governed by the migration. Finally, the strong competition phase is glassy, dominated by uninvadable state and noise-induced transitions. Our model contains, as a spatial case, the celebrated neutral island theories of Wilson-MacArthur and Hubbell. Moreover, we show that slight deviations from perfect neutrality may lead to each of the phases, as the Hubbell point appears to be  quadracritical.            
\end{abstract}
\maketitle

\section{Introduction}

Trying to characterize and quantify the factors that govern the
dynamics of natural populations, community ecologists were often
surprised by the large number of competing species that can be found
in a relatively small area. Having in mind the Darwinian picture of
natural selection and the survival of the fittest, one may expect
that a few fittest species will dominate the community (perhaps with
some sporadic presence of a  few individuals of  inferior
species) as suggested by the competitive exclusion principle
\cite{hardin1960competitive}. This is definitely not the case in
many important communities, from tropical forests to coral reef to
freshwater plankton. In fact, an understanding of the factors that
allow the maintenance of biodiversity under selective dynamics is
considered  as one of the most important challenges for modern
science \cite{science125}.

One of the versions of the biodiversity puzzle has to do with a
local community which is coupled by migration to a regional pool.
The simplest example for this scenario is the mainland-island
system, where the mainland dynamics is assumed to be relatively slow
so one can assume that the island is interacting with (i.e., receiving
immigration from)  a static pool on the  mainland. This
mainland-island model may describe any local community, provided
that the length scale involved in biological interactions (e.g.,
competition) is much smaller than the migration scale
\cite{bierregaard2010losos}.

In general, the dynamics of natural ecological communities is
subject to substantial noise. Populations are exposed to
environmental variations that affect their reproductive ability and
death rate. This effect is, typically, quite strong
\cite{tsimring2014noise,lande2003stochastic}. Even under strictly
fixed environmental conditions the stochasticity of the
birth-death-migration process (demographic stochasticity) adds
randomness to the dynamics. Under demographic noise, every finite
population goes extinct eventually, so theories of community dynamics
must include a stabilizing mechanism that makes these extinctions
extremely rare (stable coexistence) or allow for a speciation
process to maintain the species richness (unstable coexistence).

The mainland-island system incorporates features from both scenarios.
On the one hand, in a local community  there are at least a few
extinction-prone low-abundance species. One the other hand, there
are no  absorbing states in the strict sense, as individuals of any
species arrive at a fixed average rate from the mainland. Nevertheless, if
the migration is relatively weak and the local population is not
huge, some or perhaps all of the species may undergo  temporary
extinctions, leaving the island without this species until the next
individual of this species arrives from the mainland and manages to reestablish. The
statistics of these local extinction-recolonization events for birds
in North America was recently analyzed by
\cite{bertuzzo2011spatial}, below we will consider the relations
between our model and their empirical results.

Community dynamics theories are usually classified along the line
between niche and neutral. A niche theory assumes that every species
that have a non-sporadic presence on the island has its own niche.
For example, a few bird species each having a different beak size
and (correspondingly) different diet may coexist on the island if
the overlap between the niches is not too large. In the other
extreme, in a perfectly neutral dynamics there is no niche
partitioning at all, all species are using the same resources with
the same efficiency, and the dynamics is governed solely by
stochasticity. In between one can find a few "continuum models"
\cite{kadmon,gravel,pacala, zillio} that were suggested in the last
 decade and incorporate elements of neutral dynamics with
(usually weak) selective effects.

The simplest model for island dynamics is the generalized competitive Lotka-Volterra
model (GCLV) with migration.  This model is widely used in
ecology and for other applications
\cite{tilman1987importance,pigolotti2007species,fort2013statistical}.  It turns
out that the model has a very rich structure.  Our primary focus is on an individual
based 
stochastic version of the model which incorporates demographic noise.  We will see that the model, despite its simplicity, is very rich and exhibits a wide range
of different behaviors.  Our goal here is to exhibit this panoply of ``phases" and understand their origins.  A parallel
study of the much more tractable deterministic version of the model will be a key tool in
unraveling the dynamics.

In particular,
we shall show below that the most celebrated models of island
biogeography - the Wilson-MacArthur theory of island biogeography
\cite{macarthur1967theory,losos2009theory}
 and Hubbell's neutral theory of biodiversity
\cite{Hubbell2001unifiedNeutral,TREE2011} are two special cases of
this model. Following a few recent publications that  emphasized
some specific aspects of the dynamics
\cite{fisher2013niche,pigolotti2013species}, we would like to
show how the phase structure of the model is governed to a large extent by these two special limits. Finally, we will consider
the relevance of this model to the empirical findings.

\section{The stochastic GCLV model}

In this section we introduce our model and set notation.
Let us consider a regional pool of $Q$ species on the mainland. Each
individual of these species may immigrate to the island with a
certain probability per unit time, and we denote the average rate at which individuals of the $i$th
species reach the island as $\lambda_i$.

Denoting by $N_i$ the abundance of the $i$th species on the island,
the deterministic part of the dynamics satisfies
\begin{equation} \label{GLV}
\dot{N}_i = \lambda_i + \alpha_i N_i - \frac{N_i}{K_i} \left( N_i +
\sum_{j\ne i} d_{i,j} N_j \right).
\end{equation}
Here $\alpha_i$ is the growth rate of the $i$-th species on the
island, and $K_i$ sets the carrying capacity of the $i$-th species
\emph{in the absence of competition with other species}.
Interspecific interactions are expressed by the elements of the
 matrix $d_{i,j}$. Here we study a
purely competitive system where all the $d_{i,j} \ge 0$.

As our goal is to identify the different phases of this model, not
to fit it to a specific empirical system, we make a few
simplifications. First, we assume (as in
\cite{bell2000distribution}, for example) that all species have the
same flux of immigrants from the mainland, $\lambda_i = \lambda$,
the same linear growth rate that we scale to one ($\alpha_i=1$) and
the same carrying capacity $K$. We are interested here specifically in small values of $\lambda$, so that immigration primarily
serves to ``rescue" extinct species, but does not swamp the intrinsic competitive dynamics on the island.
The interaction matrix takes the
form $d_{i,j} = C c_{i,j}$, where C sets the overall strength of the
interaction and the $c_{i,j}$-s are normalized such that
\begin{equation}
\sum_{i \ne j} c_{i,j} = Q(Q-1)
\end{equation}
In other words, the average magnitude of a $c_{i,j}$ is unity. We consider here the case where the
$c_{i,j}$ are chosen randomly from a distribution with unit mean and variance $\sigma^2$; the
$c_{i,j}$ values are kept fixed throughout the process. The current work considers the case of purely competitive community without symbiosis or food web
features. Accordingly, we do consider here the case of interaction matrices with a modular or nested structure.  As we shall show be,
this structureless matrix provides a natural generalization of the Hubbell neutral theory~\cite{Hubbell2001unifiedNeutral}.  For
our simulations, the $c_{i,j}$ were chosen from a Gamma distribution with probability distribution function
\begin{equation}
P(c) = \frac{c^{1/\sigma^2-1}\sigma^{-2/\sigma^2} e^{c/\sigma^2}}{\Gamma(1/\sigma^2)}
\end{equation}
 The final form of our GCLV model is then
\begin{equation} \label{GLV1}
\dot{N}_i = \lambda +  N_i - \frac{N_i}{K} \left( N_i + C
\sum_{j\ne i} c_{i,j} N_j \right).
\end{equation}

For small immigration rate $\lambda$, extinctions and recolonizations play a crucial role in the dynamics. We treat this by constructing a stochastic individual based version of the model, so that demographic noise is explicitly included.
The number of individuals in each species is an integer.  At each time step, of duration $\Delta$, a Poisson number of immigrants of each species, with mean $\lambda \Delta$, is generated.  A Poisson number of offspring of each species, with mean $N_i \Delta$, is generated as well.  The $N_i$ veteran inhabitants
are subject to death, with the number of individuals of species $i$ that expire drawn from a binomial distribution with parameters $N_i$ and probability
$\Delta \sum_j (\delta_{i,j} + Cc_{i,j})N_j $.  Clearly, $\Delta$ needs to be chosen to be sufficiently small that this probability does not exceed unity.  The
number of individuals of species $i$ after this process is then updated to reflect the new immigrants, offspring and deaths.  On average, these changes are exactly those given by the deterministic model.

\begin{figure}
\begin{center}
\includegraphics[width=7cm]{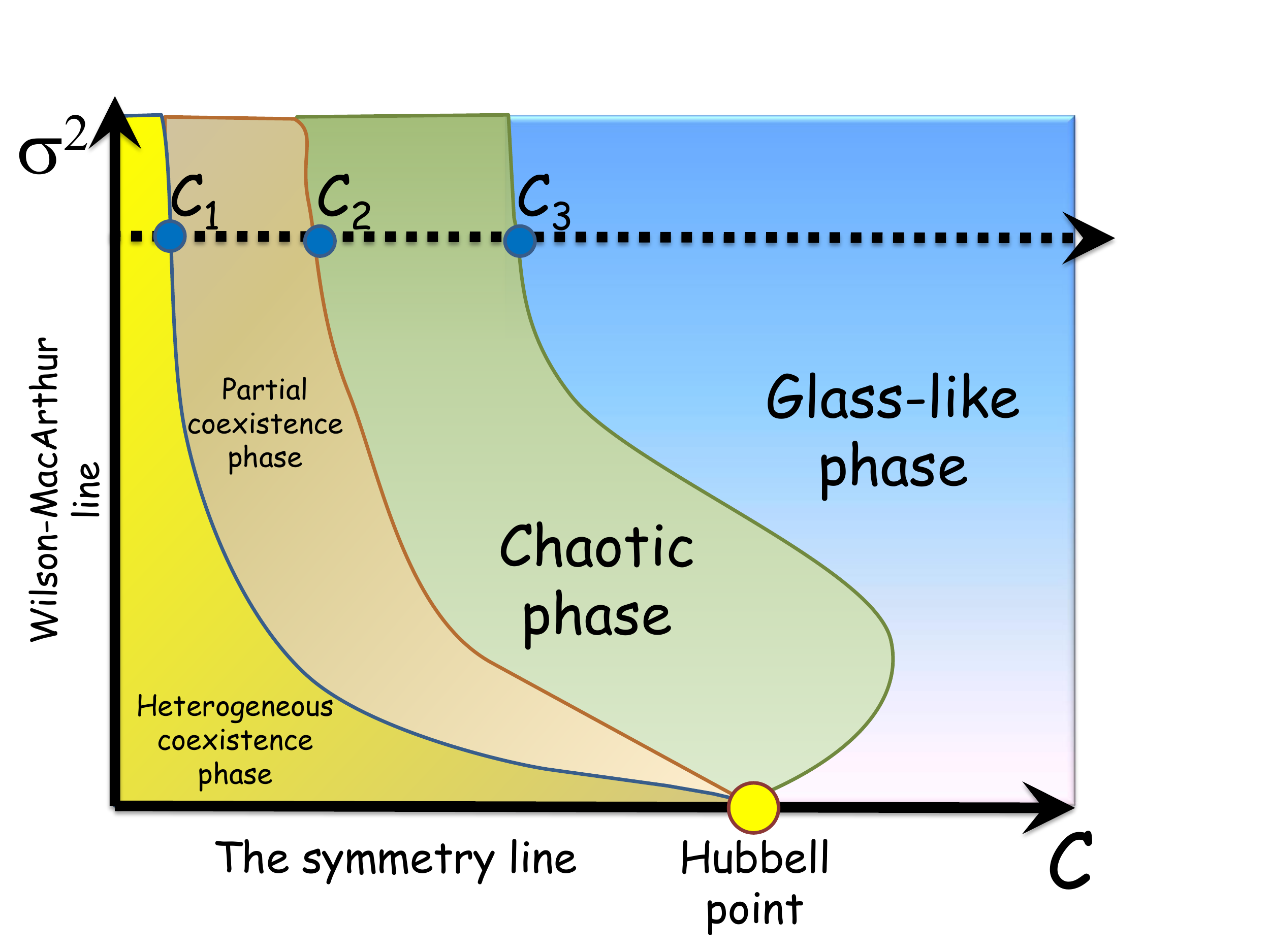}
\vspace{-0.8cm}
\end{center}
\caption{A schematic sketch of the $C-\sigma^2$ plane, showing the different phases of the generalized competitive  Lotka-Volterra system, as will be discussed below. On the  $C=0$ line    there is no competition, all species have  the same carrying capacity and the extinction-recolonization  dynamics is described by the Wilson-MacArthur theory.   At the Hubbell point, $C=1$ and $\sigma = 0$,  all individuals are equivalent and the system supports a marginally stable manifold on which the dynamics is governs only by the noise.  In the weak competition region (small $C$) all the mainland species are still semi-resident on the island, but with heterogeneous abundance and extinction times. Above $C_2$, some species are transients, with ${\cal O} \lambda$ abundance. Another increase in the strength of competition takes the system to the chaotic/intermittent phase, where the community structure changes dramatically over time and the instantaneous assembly is usually invadable. Finally in the glass-like phase  a few uninvadable equilibria control the system and the transitions are noise induced.} \label{fig1}
\end{figure}

The stochastic model is specified by four parameters: $K$,
$\lambda$, $C$ and $\sigma$. In what follows we will focus our
attention on the phases in the $\sigma-C$ plane, keeping $K$ and the
migration rate fixed. Once this behavior is
understood it is straightforward to figure out at least the qualitative features
of the dynamics for other values of migration and $K$. The different
phases in the $\sigma-C$ plane are sketched in Figure \ref{fig1}.
In the following we intend to discuss each phase in detail;
before doing that, let us focus   on two very interesting limits
that correspond to the $x$ and the $y$ axes of Figure \ref{fig1}.

\section{The Wilson-MacArthur line, $C=0$}

 When
$C=0$, species do not interact with each other and the dynamics of
each species is logistic with carrying capacity $K$.
Deterministically, an infinitesimally small population grows
exponentially and then saturates to the steady state value, $K$. With
demographic noise the first immigrant may fail to establish a
community (e.g., it may die before reproduction, even if the birth
rate is larger than the death rate), so not every immigration
results in successful colonization. The chance of a successful colonization depends on
the details of the stochastic process (in particular, the variance
in the number of offspring per individual, see the analysis of
\cite{DAKSIS} for the SIS model). After a successful recolonization the
population still fluctuates around $K$, and in the long run it must
go extinct as well since (without migration) the zero population is
an absorbing state. The rate of these long-term extinctions (as
opposed to colonization failures that take place at short times)
depends, again, on the details of the process and the value of $K$
\cite{OIK19991}.

Accordingly, in the $C=0$ limit of our model the history of every
species is made of a series of local extinctions and recolonization
events, and the rates of extinction and recolonization are equal for
all species. What emerges from this scenario is the celebrated
Wilson-MacArthur model of island biogeography: the species richness
on the island, $S$, satisfies $\dot{S} = -e S + r(Q-S)$, where $e$
is the extinction rate and $r$ is the recolonization rate, both
rates will depend on $K$ and on the details of the stochastic
process. The Wilson-MacArthur prediction for the  average species
richness is $\bar{S} = rQ/(r+e)$, the typical size of species
richness fluctuations is $\sqrt{S}$,  and the statistics of sojourn
times (the periods between colonization and extinction, as well as
the periods between extinction and colonization) is exponential.

When the only stochastic effect taken into account is  demographic
noise, as is the case in this paper, the chance of extinction
decreases exponentially with $K$  \cite{kessler2007extinction} and therefore for the
typical values of $K$ considered here $e \ll 1$. If this is actually
the case, or when $\lambda$ is large so $r \to \infty$, all mainland
species are presented on the mainland up to tiny short-term
fluctuations.($S \approx Q$). Accordingly, in our model one observes
Wilson-MacArthur dynamics on reasonable time scales only when $K$ is relatively small.  
More realistic models have to take into account other types of noise
(including environmental variations, attacks by pathogens etc.) that
may lead to extinction, and the Wilson-MacArthur model will then be
relevant even for higher values of $K$.

\section{The symmetry line and the Hubbell point}

On the $x$ axis, the  $\sigma = 0$ line, species do interact with
each other but the interaction is symmetric; i.e., no change in
community dynamics occurs upon switching the species labels of any
two given populations. $C$ measures the strength of interspecific
competition: if $C<1$, the intraspecific competition is stronger
than the interspecific (reflecting mechanisms like resource
partitioning or frequency dependent predation) and a low-density
species may invade the system. On the other hand, for any $C>1$ the
intraspecific competition is weaker than the interspecific, resulting in
competitive exclusion \cite{chesson2000mechanisms}. At the boundary
between these regimes one finds the Hubbell point $C=1$ (see Figure
\ref{fig1}). At the Hubbell point the model is \emph{neutral}: any
individual competes equally with any other individual and the strength of
each pair competition is fixed and independent of  species
affiliation.

Without immigration, the deterministic GCLV supports, to the left of
the Hubbell point, an egalitarian coexistence stable fixed point
where all species are present on the island with the same
abundance $K/(1+(Q-1)C)$, while above the Hubbell point the stable
solution admits only one species with abundance $K$, all other
species has zero abundance  (the identity of the surviving species
is determined by  initial conditions). At the Hubbell point the
deterministic dynamics supports a marginal manifold: every combination of $N_i$'s
such that the total population is $K$ is a solution of Eq. \ref{GLV1}. The simplicity of the GCLV at the Hubbell point
allows one to solve analytically for the species abundance
distributions (SAD) even with demographic noise \cite{maritan1},
environmental stochasticity \cite{malcai2002theoretical}, or a
combination of demographic and environmental noise
\cite{kessler2014neutral}.

About 15 years ago, Hubbell \cite{Hubbell2001unifiedNeutral}  put
forward his very influential neutral theory of biodiversity,
suggesting that all species and all individuals are demographically
equivalent and the only mechanism that drives the system is pure
demographic noise and the (typically slow) rate in which new species
are introduced. Hubbell's model has two versions. In the
metacommunity version speciation is the mechanism that leads to the
introduction of a new species, while in the mainland-island version
colonizers of new species arrive from the mainland (assuming large $Q$). Some
patterns predicted by this neutral theory, and in particular the
species abundance distribution (SAD) on the island, fit quite nicely
those recorded in many empirical studies. 

Off the Hubbell point the infinite degeneracy is lifted, but the
analysis of the stochastic system is still relatively easy since
every species may be analyzed independently, the effect of all
individuals from  other species may be encapsulated into a single
parameter. This feature was exploited by \cite{pigolotti2013species}
who manage to solve the (metacommunity version) stochastic GCLV
analytically along the $\sigma = 0$ line. As expected, below the
Hubbell point the SAD shows a peak around the deterministic value
$K/(1+(Q-1)C)$, and the sharpness of this peak increases as $C$
decreases.

\section{An overview of the dynamics}
\begin{figure*}
\includegraphics[width=0.85\textwidth]{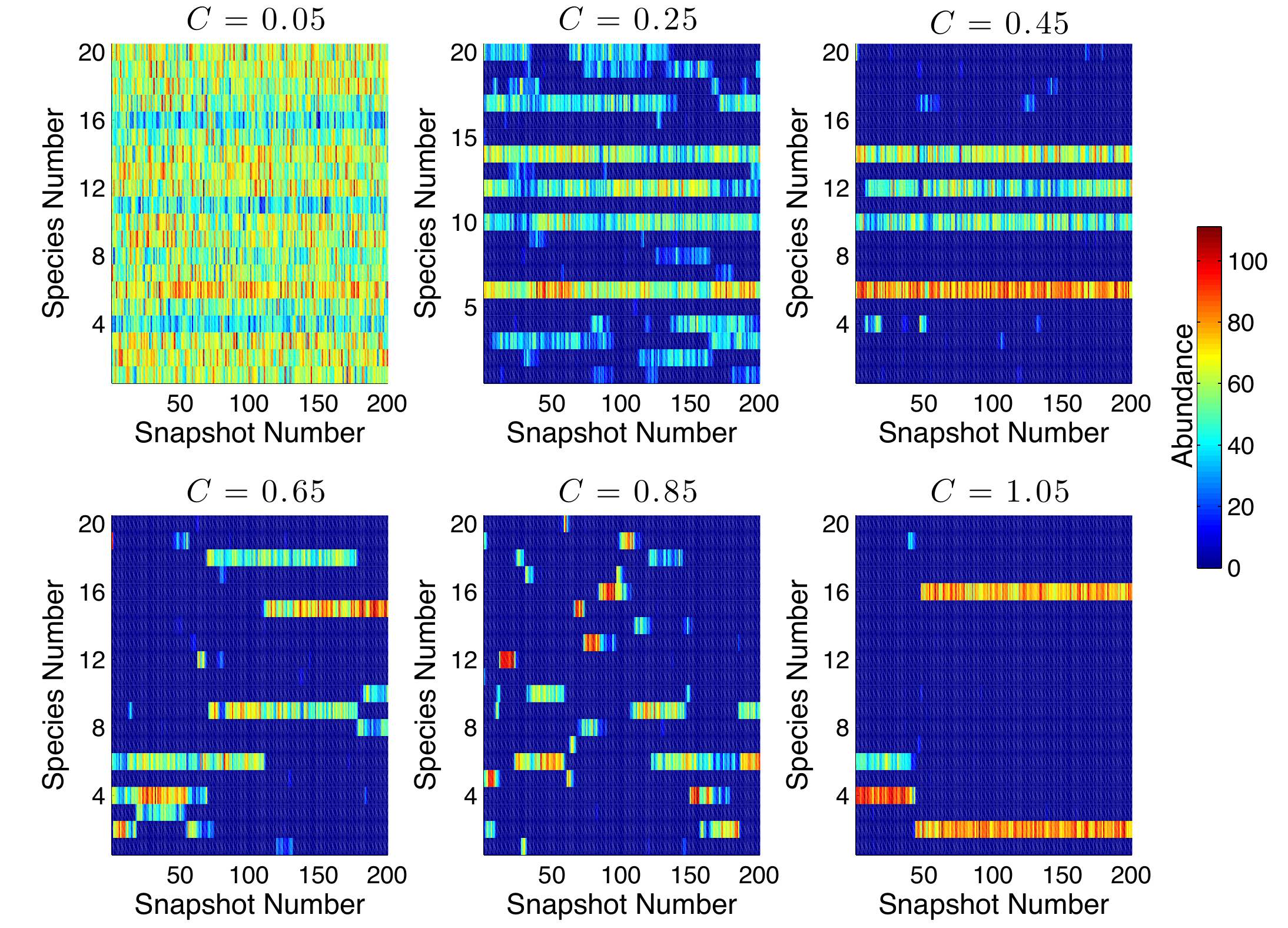}
\caption{Snapshots of the abundance for all species in the stochastic model for various levels of competition, $C$.   Red is high abundance, dark blue low, as shown in the
color bar at the right.  $Q=20$, $K=100$,
 $\sigma=0.5$, $\lambda=0.01$. The time between snapshots is $\lambda t=0.08$.} \label{fig all}
\end{figure*}

We present in Fig. \ref{fig all}  sample runs of the system for $K=100$, $\lambda=0.01$, $\sigma=0.5$, $Q=20$, for varying $C$.  The abundance of each species is indicated by color.  The horizontal axis is time, with snapshots of the system taken every $t=0.08/\lambda$.  The vertical axis is species number, and runs from 1 to $Q=20$.  It is clear that the system exhibits very different behavior as $C$ is varied.  One can identify four different generic  behaviors, which we shall call ``phases".  Very briefly, in the first, low $C$, phase all $Q$ species are present essentially all the time.  In the second, some species are basically no longer present, supported only by the infrequent stochastic immigration of new individuals, while the rest have a more or less continuous existence, with occasional temporary extinctions. The third phase at yet higher $C$ is the most complicated, with the system jumping from one state to another.  These states would be stable in the absence of immigration, but a successful colonization by some specific species drives the system to a new quasi-stable state.  In the last phase, the system spends a preponderance of time in some stable state. In the following, we will attempt to explicate in more detail the various features of each of these phases.

\begin{figure}[b]
\includegraphics[width=0.45\textwidth]{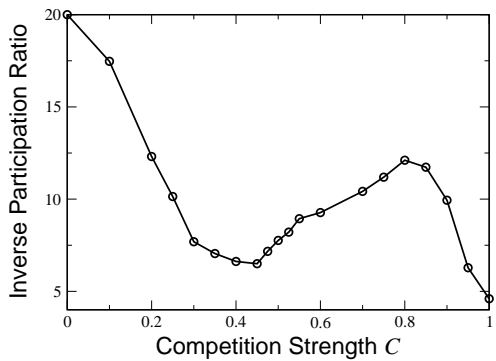}
\caption{The inverse participation ration, $\textrm{IPR}$, as a function of $C$. $N=20$,  $K=100$, $\sigma=0.5$, $\lambda=0.01$.}
\label{figIPR}
\end{figure}

One way to quantify the different behaviors is via the ``inverse participation ratio", IPR, defined as
\begin{equation}
\textrm{IPR} \equiv \frac{\left(\sum_{i=1}^Q \langle N_i  \rangle \right)^2 }{\sum_{i=1}^Q \langle N_i \rangle ^2}
\end{equation}
The angle brackets refer to an average over time.  The IPR varies from 1 to $Q$.  In the case where only one species is present, it takes the value unity, and if all species have equal abundance, its value is $Q$.  Thus, the IPR is a measure of how many different species are active in the system.  It can differ enormously from the time average of the instantaneous number of species present.  We show in Fig. \ref{figIPR} the IPR as a function of $C$ for runs with the same parameters as in Fig. \ref{fig all}.  We see that the IPR is not monotonic in $C$.  It initially decreases from $Q$, reaches a minimum and then starts to increase.
It then achieves a local maximum and then starts to decrease again.  This change in behavior of the IPR is clearly reflective of the different patterns captured in Fig. \ref{fig all}.  We shall elaborate on the behavior of the IPR as we investigate each phase.

\section{Phase I: The heterogeneous  coexistence phase}

We start by considering  the leftmost region of Fig. 1, the area where $C$
is relatively small. Increasing  $C$ from zero,  at a fixed value
of $\sigma$, corresponds to an increase of interspecific competition;
for example, different species of birds that leave happily together,
each having a different diet of worms, may start to interact with each
other if the supply of worms is decreased and different species
begin to consume the same resource. In such a case species-specific
niches are ``squeezed" towards each other, increasing the niche
overlap.

Once the species start to compete, the heterogeneity of the
$c_{i,j}$'s (as reflected in the parameter $\sigma$) implies that
some species are impacted by the competition more than the others.
As a result, the abundance $N_i$ of the $i$th species  is
no longer $K$  (as  on the Wilson-MacArthur line
$C=0$); instead, all abundances are reduced by a species-dependent amount
and one gets a distribution of species abundance
values. Still, as long as $C$ is not too large, the deterministic
dynamics of Eq. (\ref{GLV}) supports a single attractive fixed point
that corresponds to the case where all the $Q$ species of the
mainland are represented on the island. In this case, we can approximate the
system by the $\lambda=0$ system, since the small external flux does not qualitatively change the
steady-state, which in any case has all species present. Eq. \ref{GLV} with $\lambda
= 0$ allows for an explicit stationary state solution,
\begin{equation}
N_i = K B^{-1} \mathbb{1}
 \label{steady}
 \end{equation}
where $\mathbb{1}$ stands for the length $Q$ column vector
consisting of all one's, and  $B$ is the matrix
\begin{equation}
B_{i,j}=\delta_{i,j} + C c_{i,j}.
\end{equation}

\begin{figure}
\begin{center}
\includegraphics[width=0.45\textwidth]{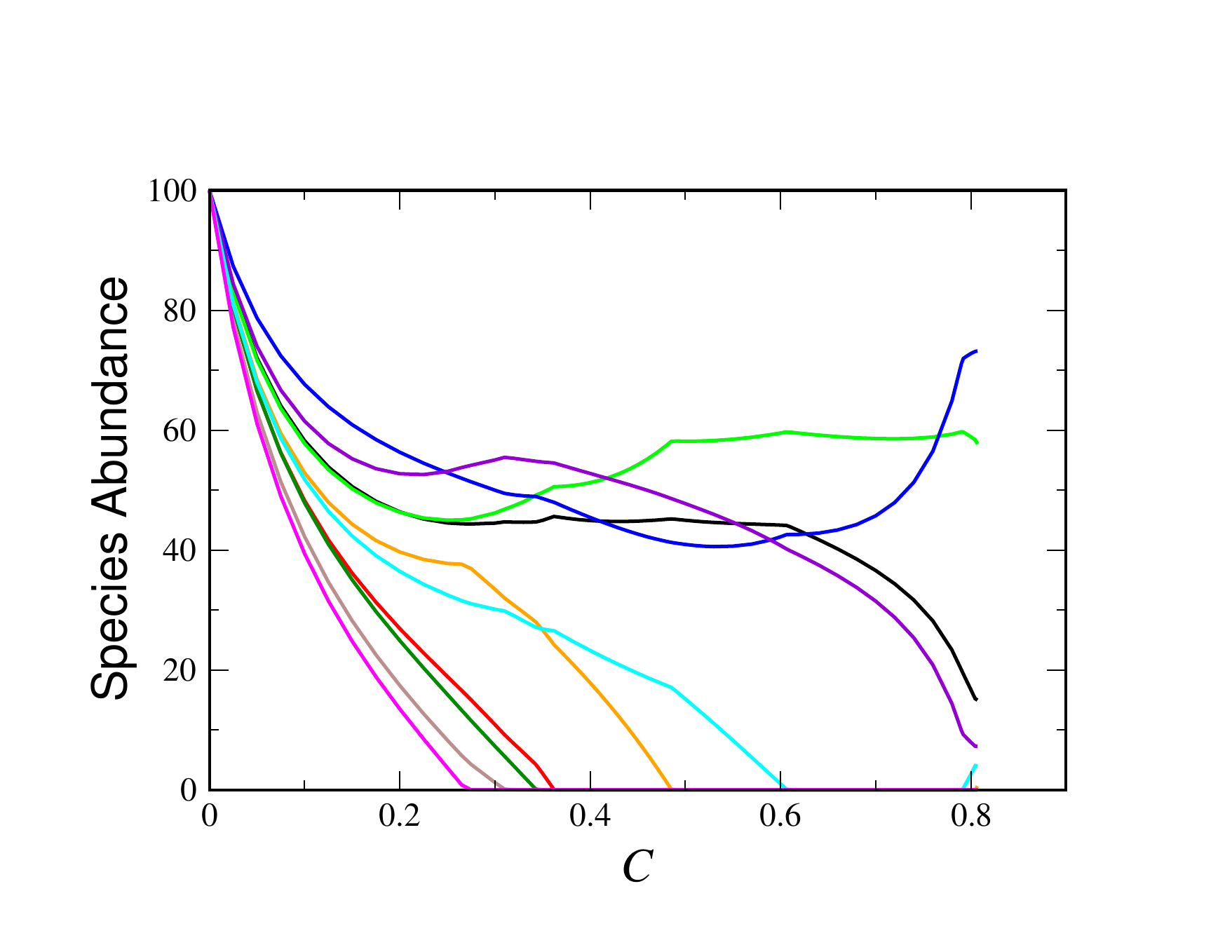}
\vspace{-0.8cm}
\end{center}
\caption{\textbf{Abundance vs. competition}. The abundance of $Q=10$
species on the island is plotted vs. the competition parameter $C$,
for a deterministic  GCLV model with $K=100$ and $\sigma = 0.5$,
without immigration ($\lambda=0$). In the non-interacting limit  $C=0$ the
abundance of all species is identical and equal to $K$. At $C_1
\approx 0.28$ the abundance of the ``weakest" species reaches zero -
this is the May transition, where the system enters the competitive
exclusion phase. Upon increasing $C$, more and more species go
extinct until (here, around $C=0.78$) a species comes back to life,
as the competitors that suppressed it were themselves suppressed. }
\label{fig2}
\end{figure}

Figure \ref{fig2} illustrates the process. For $C$ small enough that
coexistence fixed point solution is physical, such that all $N_i$ are positive, there is a unique stable solution and 
all the $Q$ mainland species will
be present on the island. Their abundances decrease as a function of
$C$, but  no species goes extinct in the deterministic theory.
As long as this remains true, for reasonably large $K$, one can more or less neglect  the noise, given the stability of the fixed point solution.
 The species that have
lower abundance are those who suffer more from the competition, and
even if demographic noise drives a few of them to extinction, the
effect on the rest of the network is minor and the system will
restore itself by immigration.  This situation is illustrated by the first panel of Fig. \ref{fig all}, the case $C=0.05$, where
 there are no extinctions seen in the time-frame shown.  Examined over a much longer period, there are
indeed a few occasional extinctions of all the various species.  The most extinction prone species, for example, was seen to
go extinct a total of four times over a period of $24,000$ snapshots.

In such a scenario one expects deviations from  the Wilson-MacArthur
formula. When the abundance of the species are different from each other, the chance of
extinction $e$ becomes species dependent, so once $C>0$ the sojourn
times will reflect a convolution of exponents with different
timescales. When the community heterogeneity increases,  the Wilson-MacArthur
extinction-recolonization dynamics is most relevant for the smaller species, again with heterogeneous
statistics of extinction times. 

\section{The May transition and Phase II: the partial coexistence phase}

As $C$ increases even more, the deterministic dynamics (without
immigration) no longer supports a steady state with
 all the $Q$ mainland species coexisting. In the heterogeneous
coexistence phase, as discussed above, the $\lambda=0$ steady-state system admits a solution, Eq. (\ref{steady}), which is
both feasible ($N_i >0$) and stable (the real part of all the
eigenvalues of the community matrix are negative). As follows from the work of
May \cite{may1972will}, 
the chance for the system
to fulfill these requisites, for fixed heterogeneity $\sigma$, decreases exponentially with $Q$. In
fact, for the GCLV in the coexistence phase, the main obstacle is
feasibility \cite{rozdilsky2001complexity}.  Eq. (\ref{steady})
suggests that the solution will be feasible when the sum of all the
rows of $B^{-1}$ is positive. As the average sum of a row approaches
zero with $C$, the chance to pick only positive values for $N_i$
decreases exponentially with $Q$.

In Figure \ref{fig2} one observes that, for the particular realization of the $c_{i,j}$ being simulated here, with $Q=10$,  the abundance
of the weakest species reaches zero around $C=0.28$. This point marks the transition
from the heterogeneous coexistence phase to the partial coexistence
phase. In this phase the island species richness, $S$, is smaller
that the mainland richness $Q$, as some populations are not
supported anymore on the island. Weak competitors (species that
suffer from strong competition against others that do not suffer as
much) are selected out.

This is the deterministic picture without migration. With migration,
``exclusion" does not correspond to exactly zero density. Instead,  above the May
transition the deterministic density of these species is
$\cal{O}(\lambda)$. Since we assume that $\lambda$ is small,
demographic noise induces frequent extinction of these transient
species. As opposed to the other, ``semi-resident" species (the ``semi-" prefix taking account
of the possibility of a short-lived absence due to demographic fluctuations), the growth rate
of a transient species immigrant is negative, rendering its average
persistence times small, independent of $K$. 

Technically speaking, the May transition takes place at a  critical
value of the competition parameter  $C_1$,  where  the smallest
$N_i$ reaches zero. Above the transition  one can define a reduced
system, eliminating the row and column of $B$ corresponding to the
eliminated species. This reduced system does admit a solution where
all the (remaining) species have positive abundance.  This is clear,
since at the exact value of $C$ at which the next species vanishes,
the remaining $N_i$'s constitute a solution of the reduced problem.
In addition, the equation for the species with vanishing abundance (call it $k$)  reads
\begin{equation}
0=1  - (C/K)\sum_{j \ne k} c_{k,j} N_j
\end{equation}
The r.h.s. of this equation is just the growth rate of the $k$th
species, so that for $C$ smaller than the critical value, the growth
rate of this species is positive, and for $C$ above this value, it
is negative.  Thus, the $k$th species cannot invade for $C$'s
slightly larger than the value at which that species disappears from
the community.  

\begin{figure}[t]
\includegraphics[width=0.45\textwidth]{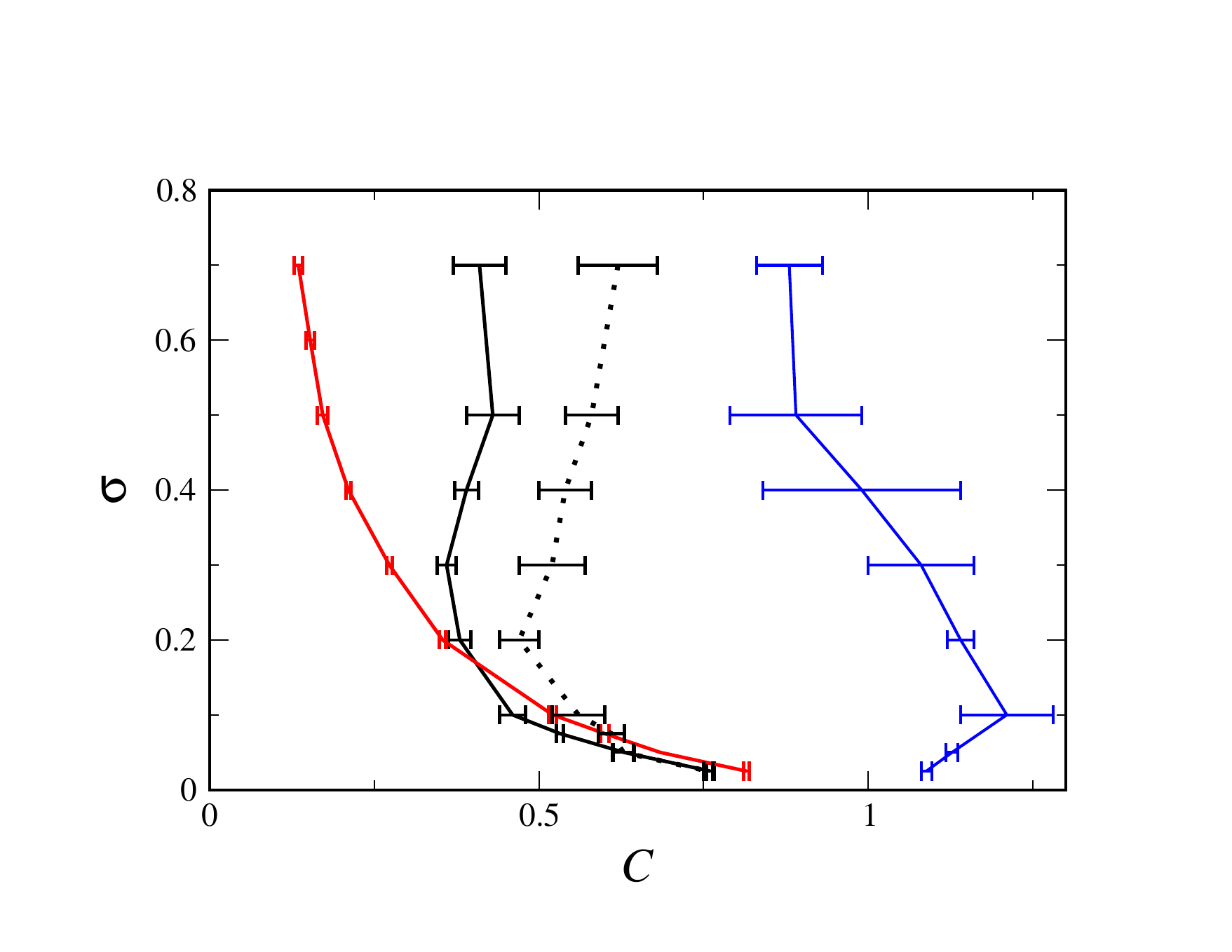}
\caption{The phase boundaries in the $C$, $\sigma$ plane. The leftmost, red, line is the May line where the deterministic solution with all species present
disappears. The next, solid black, line to the right is the simulationally determined line where the concave downward part of the species occupancy curve disappears, for the
case $K=100$, $\lambda=0.01$.  The  dotted black line to the right of the solid black line is the same line for $K=200$, $\lambda=0.01$. The rightmost, blue, line marks the region where the system spends more than $20\%$ of its time in a particular state, for $K=100$, $\lambda=0.01$. The May line was determined by averaging over 100 different realizations of the interaction matrix.  The other lines result from averages over 10 or 20 matrices.}
\label{phase}
\end{figure}

In Fig. \ref{phase}, we present the May line in the $C$-$\sigma$ plane, showing the dependence of $C_1$ on $\sigma$.  The data was obtained by
calculating the May point for a set of 100 random $c_{i,j}$ matrices for a given sigma, and averaging.  We see that $C_1$ increases with decreasing 
$\sigma$, and appears to approach unity for $\sigma \to 0$.  We shall return to this point later.

Increasing $C$ in this manner in the partial coexistence
 phase, one obtains a nested hierarchy of solutions, each
with less diversity, which are immune to invasion by any of the
extinct species. As mentioned, this hierarchy is completely independent
of the carrying capacity $K$.
This prescription eventually breaks down.  At some point,
there is a ``resurrection" of one of the eliminated species. This
happens due to the fact that the species in question was strongly
suppressed by another species. As $C$ increases, this suppressor
species is itself reduced in abundance, and so the suppressed
species is able to stage a comeback.  Thus there is a value of $C$
at which this suppressed species is able to invade.  At this point,
it needs to be added back to the reduced system, since the small external flux will
reintroduce it and it will then grow in abundance. Increasing $C$
further, things continue on in this fashion, losing and regaining
species. The main characteristic of the partial exclusion
phase, the distinction between semi-resident
species that admit a finite population  and $\cal{O}(\lambda)$
transients that cannot invade,  still holds.

\begin{figure}[t]
\includegraphics[width=0.45\textwidth]{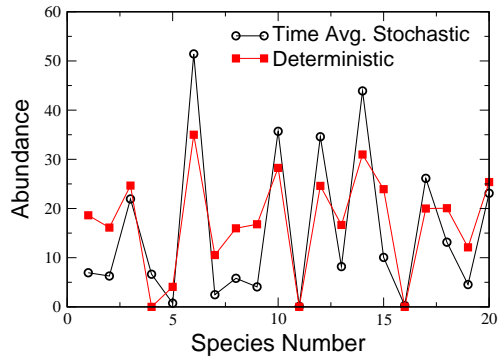}
\caption{The time averaged abundance (open circles) vs. species number for an extended run of the second panel of Fig. \ref{fig all}. $C=0.25$. $N=20$,  $K=100$, $\sigma=0.5$, $\lambda=0.01$.  Also shown is the deterministic time-independent solution (filled squares).  The lines are shown just to guide the eye.}
\label{figdetvsmean}
\end{figure}

An example of the stochastic version of this partial coexistence phase can be seen in the
second and third panels of Fig. \ref{fig all}.  For this realization of the $c_{i,j}$, with $Q=20$, the May transition point is at approximately
$C_1 \approx 0.169$.   We show in Fig. \ref{figdetvsmean} the time averaged abundances of the species for an extended version of the run presented in the second panel of Fig. \ref{fig all}, with $C=0.25$, past the May point.  We see that the deterministic solution is essentially missing 3 species, $\{4,\,11,\, 16\}$.  The stochastic run shows that the overall structure of the abundance vs. species of the deterministic model is preserved.  The smaller species are, as might be
expected, more severely impacted by the demographic noise,  to the benefit of the larger species, which do not have to suffer as much competition from these small species.  As $C$ increases, not only does the number of species in the deterministic solution decrease, but a significant fraction of these have very small abundances. The upshot is that the number of species in the stochastic simulation with significant abundance decreases
significantly.  This is reflected in the sharp fall of the IPR in this range of $C$.

\section{Phase III: The ``Chaotic" Phase}

\begin{figure}[t]
\includegraphics[width=0.45\textwidth]{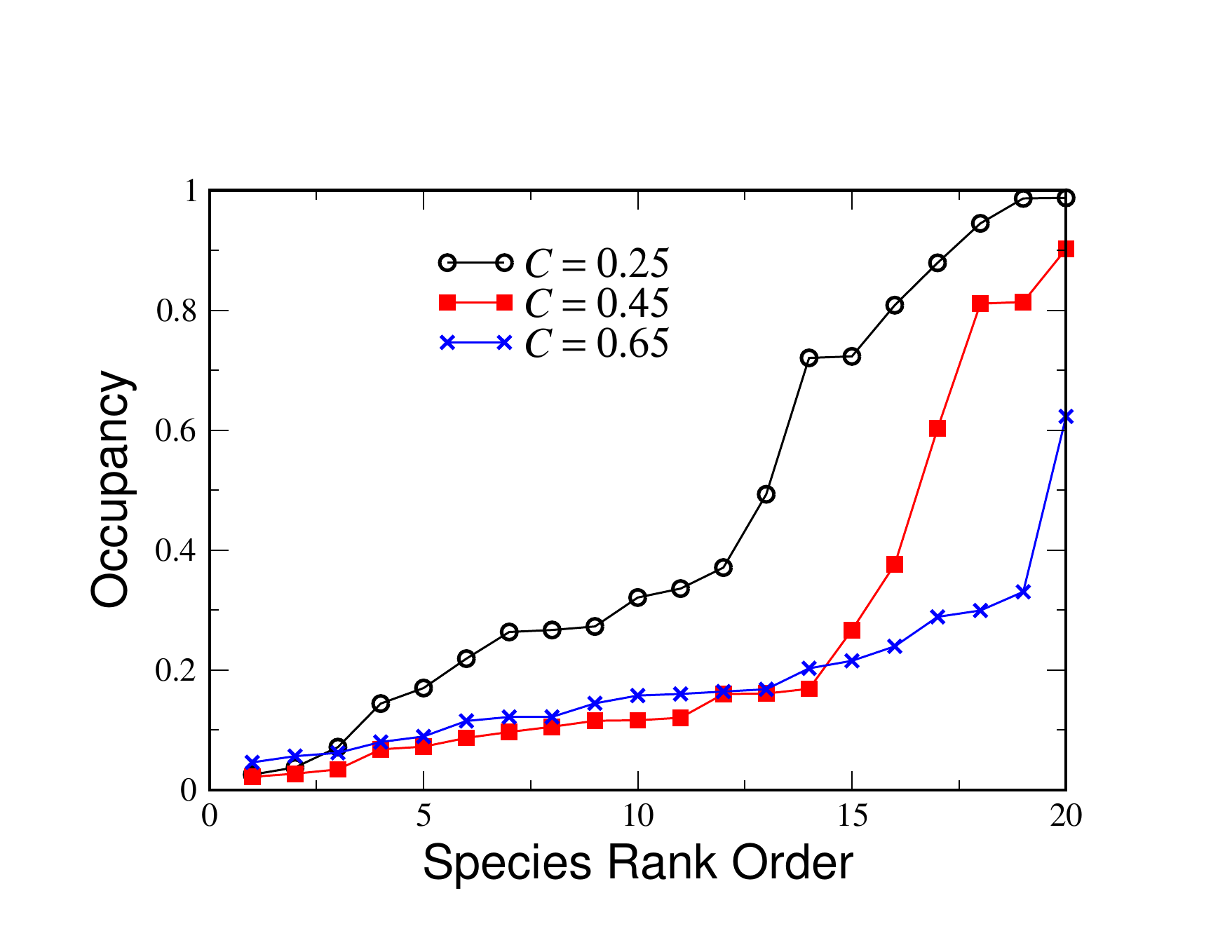}
\caption{The occupancy vs. rank species order (of increasing occupancy) for an extended run of the second, third and fourth panels of Fig. \ref{fig all}. $C=0.25$ (open circles), $C=0.45$ (filled squares), $C=0.65$ ($\scriptstyle{\textrm{x}}$'s). $N=20$,  $K=100$, $\sigma=0.5$, $\lambda=0.01$.  The lines are shown just to guide the eye.}
\label{figocc}
\end{figure}

At some point, $C_{2}$, the partial exclusion picture breaks
down. This is clear from an examination of the fourth and fifth panels of Fig. \ref{fig all}.  There is no longer a fixed set of ``resident" species which are always present in large numbers. Instead, there is a constant turnover in the set of species present.  This is reflected in the rise seen in the IPR in Fig.
\ref{figIPR} for $C\gtrsim 0.45$.  Another way to see the same point is to examine what we will term the ``occupancy" of each species, the fraction of time that it is present on the island.  In Fig. \ref{figocc}, we show the occupancy for the (extensions of) the runs in Fig. \ref{fig all}.  We see that for $C=0.25$,
in the partial coexistence phase, there is a set of seven species with relatively high occupancy.  These give the overall curve for $C=0.25$ a concave
downward form at high rank order. The number of high occupancy species drops to three at $C=0.45$, and at $C=0.65$ the concave downward part of the curve has disappeared and the entire curve is convex upward.  Thus, beyond the rise in the IPR, another sign of the end of the partial coexistence phase is the disappearance of the
concave downward part of the occupancy curve.  We will adopt this as our operational criterion for the location of the phase boundary.  It turns out that this is a more robust measure than the IPR curve, since the latter requires very long runs to measure accurately and the existence of a local maximum is
partially masked by intrinsic small scale oscillations in the IPR curve due to the discrete nature of the problem.  In practice, what we do is construct the straight line curve connecting the first and last points of the occupancy curve.  If any point in the right half of the occupancy curve lies above this line, the
value of $C$ is assigned to the partial coexistence phase.  In Fig. \ref{phase}, we show the phase boundary measured in this manner.  We find that
the partial coexistence phase becomes narrower as $\sigma$ decreases, and disappears below some value of $\sigma$.

It is hard to identify a single clear criterion from the deterministic model that controls the transition out of the partial coexistence phase to this ``chaotic" phase. In some cases the system has no stable stationary solution for a range of $C$, in which case we have clearly the partial coexistence phase.
In other cases, there are one or more stable stationary solutions, but with  small basins of attraction such that generic initial conditions do not flow to them.
In the stochastic model,
the transition as measured according to our operational definition clearly depends on the carrying capacity $K$.  Increasing $K$ moves the start of the
``chaotic" phase to larger $C$, as can be seen in Fig. \ref{phase}.

\begin{figure*}
\includegraphics[width=0.45\textwidth]{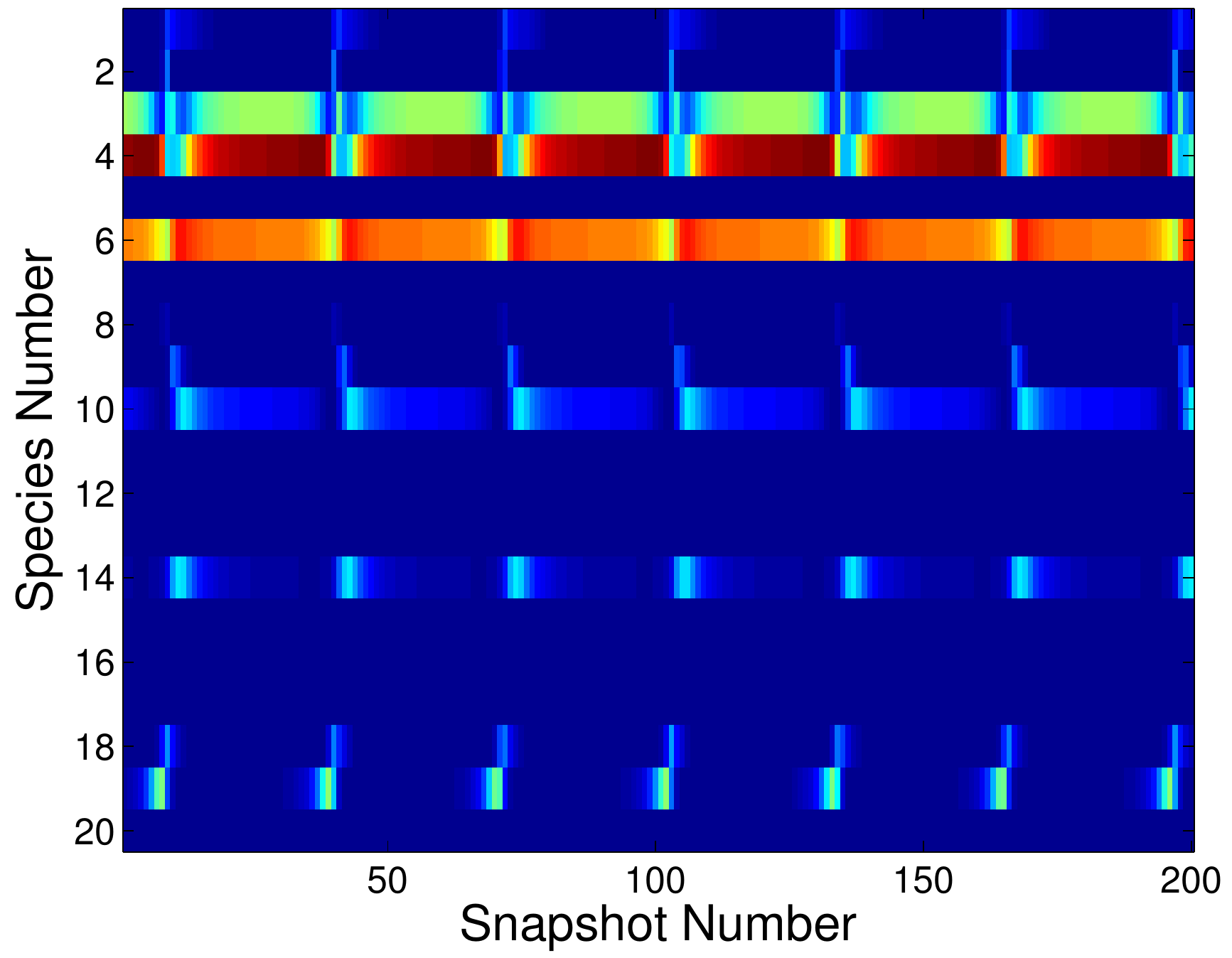}
\includegraphics[width=0.45\textwidth]{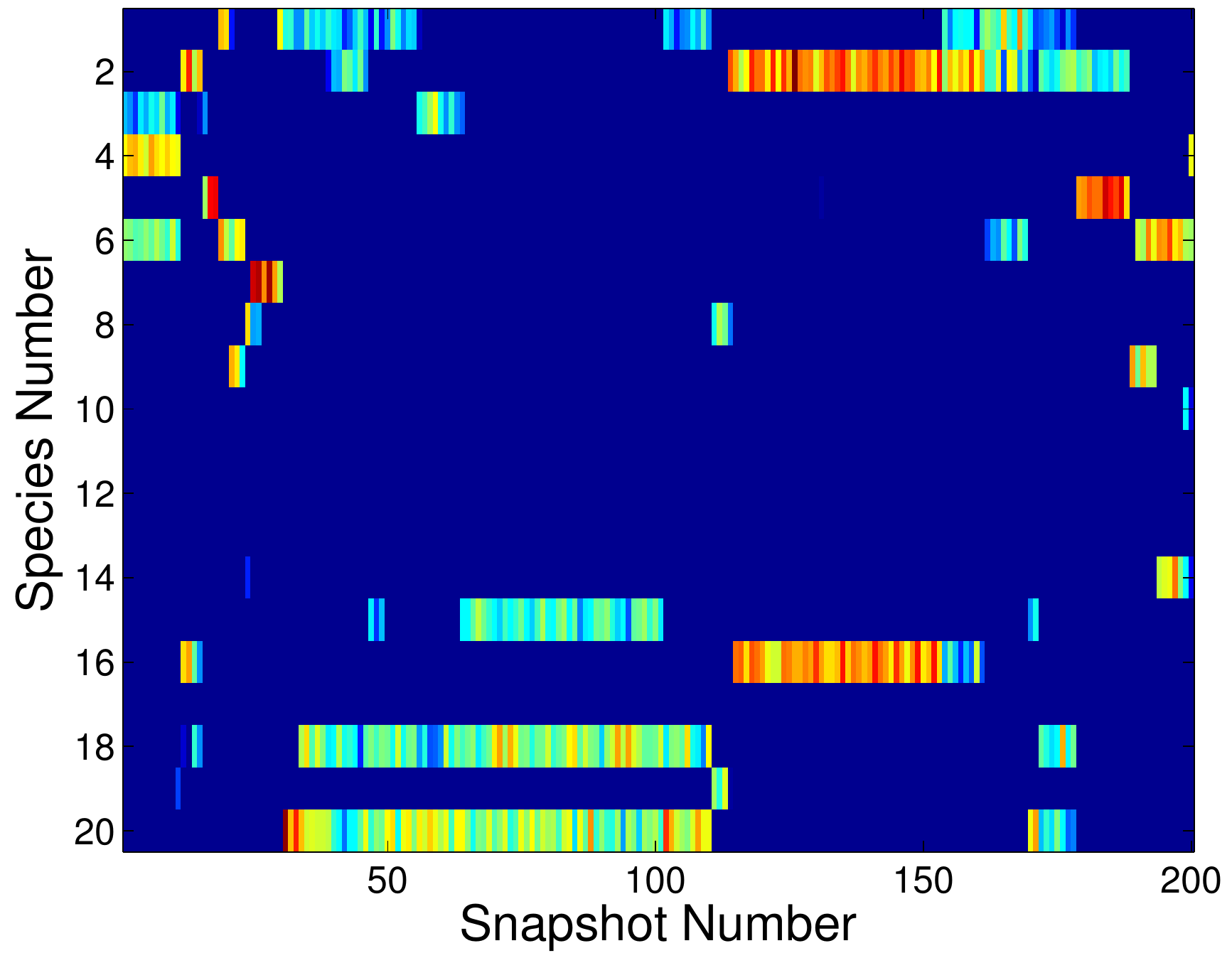}
\caption{Snapshots of the abundance for all species.   Red is high abundance, dark blue low.  $Q=20$, $K=100$,
$C=0.675$, $\sigma=0.5$, $\lambda=0.001$. Left: The
deterministic model, with time between snapshots of  $\lambda
t=0.016$. Right: The stochastic model, with time between snapshots of $\lambda t=0.128$. This larger period was chosen
in order to exhibit the wide variety of states generated by the stochastic model.} \label{figosc}
\end{figure*}

The stochastic system in the ``chaotic" phase shows a
very interesting behavior that differs substantially from the
predictions of the deterministic equations. Instead of  smooth orbits the system appears to
visit a state for a relatively long time, leave it and enters into a
period of mess, gets stuck again into another state for some period and
so on.  An example of this can be seen in Fig. \ref{figosc}, where the deterministic behavior in the left panel
stands is sharp contrast to that of the stochastic system with $K=100$ in the right panel.

\begin{figure}
\includegraphics[width=.45\textwidth]{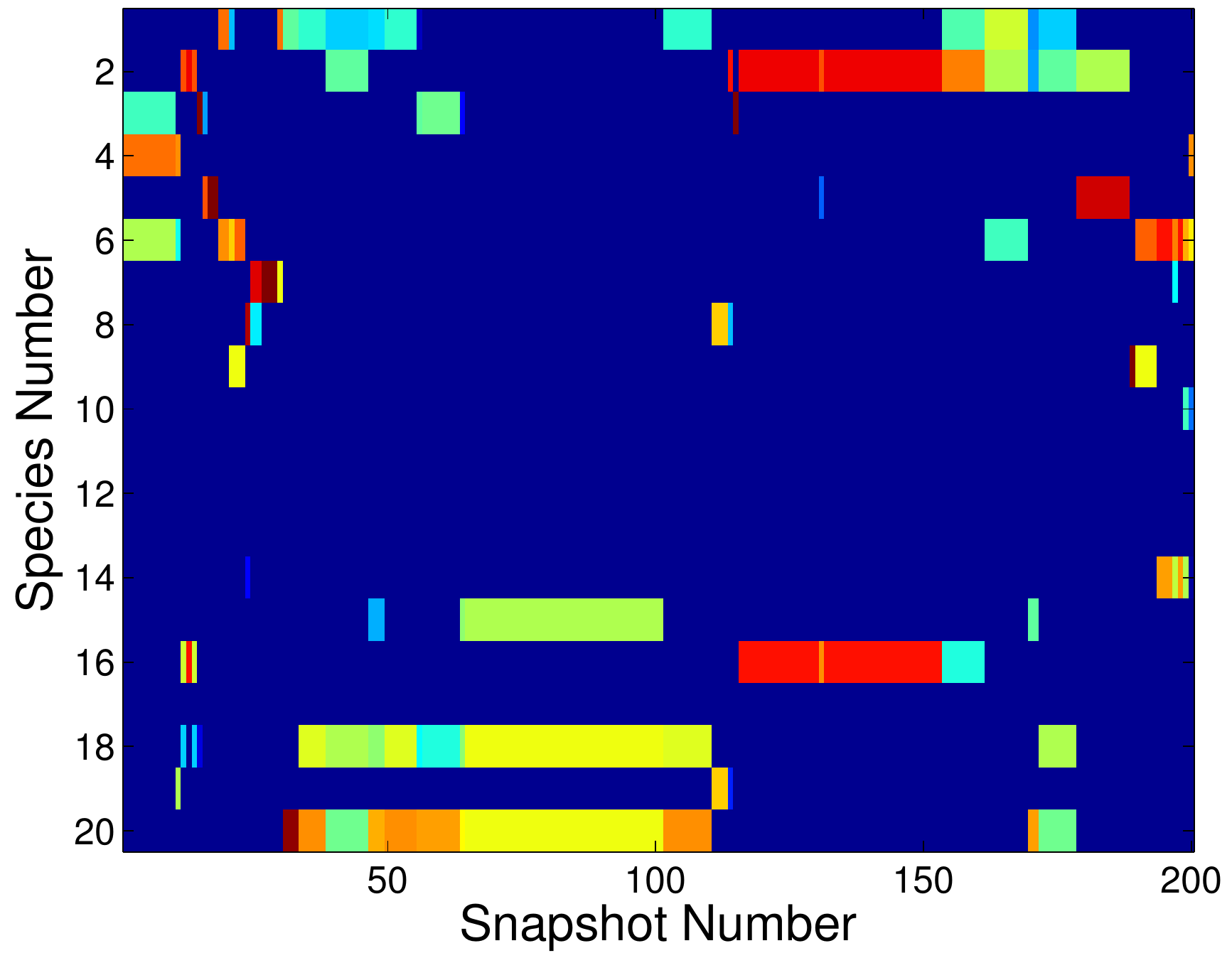}
\caption{Snapshots of species composition of the local attractor, as
obtained from the right panel of  Fig \ref{figosc} by the snapshot quenching technique. }
\label{figquench}
\end{figure}

To clarify the behavior of the stochastic system, we have
implemented a snapshot quenching technique, where we attempt to  filter out the less interesting small fluctuations within a 
given basin of attraction while preserving the major qualitative regime shifts between different basins. Taking the state of the
stochastic system at a certain time $t$, we have used it as an
initial condition for the deterministic dynamics of Eq. (\ref{GLV1})
with $\lambda = 0$, and integrate numerically this system until it
relaxed to a steady state. (Very rarely, the system relaxes to a more complicated dynamical state, which we
characterize by its species content.) This state supports, typically, only a
few species, where all others are absent either because of the
initial conditions (note that the demographic noise leads to local
extinctions) or because of the selection pressure in the
deterministic dynamics. This snapshot
quenching procedure indicates the species composition of the local
attractor at $t$. Repeating this procedure at $t+\Delta t$, the
quenching may relax to the same configuration (if the system remains
in the basin of attraction of the same attractor) or to another
configuration. Figure \ref{figquench} shows the results of the quenching
procedure for the same stochastic dynamics depicted in the right panel of Fig.
\ref{figosc}.

\begin{figure}
\includegraphics[width=0.45\textwidth]{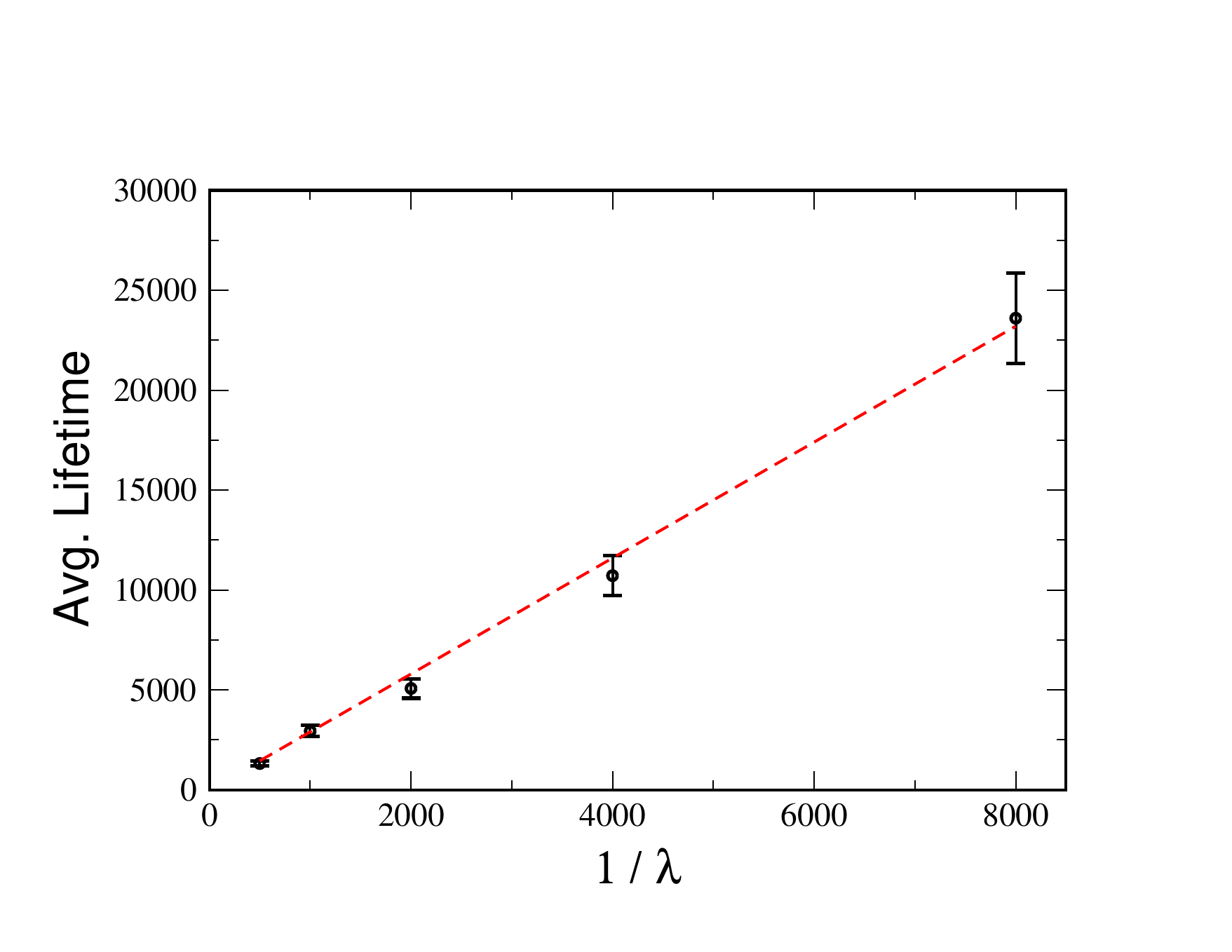}
\caption{ The average lifetime of the invadable state with species $\{15,18,20\}$ for the $Q=20$, $C=0.675$ system
simulated in Fig. 4, as a function of $1/\lambda$, showing that the lifetime is inversely proportional to $\lambda$. The dotted curve is a straight line
to help guide the eye.}
\label{fig6}
\end{figure}

The dramatic difference between the deterministic and the stochastic
dynamics in this phase has to do with local extinctions. Under
demographic noise a species with a deterministic orbit that takes it
close enough to zero abundance  may go extinct, and once this
happens, it goes out of the game until the  next immigrant from this
species arrives from the mainland and manages to establish. With
small values of $\lambda$, this quasi-stable state persists for
relatively long times, many species are absent from  the
competition so the effective number of species is smaller and the
system can find a steady state, as if it was in the competitive
exclusion phase.

However, unlike the local minima in a glassy energy landscape, here
every local attractor is unstable to invasion by at least one of the $Q$ species
on the mainland. The dwell time distribution for an invadable state is exponential,
with the mean dwell time inversely proportional to  the migration rate $\lambda$ (at least for not too small $K$).
This relationship is seen clearly  in Figure \ref{fig6}, where the mean dwell time is plotted against $1/\lambda$ for a particular invadable state.
 Accordingly, the
stability of a local attractor is determined by two factors. One is
the number of potential invaders, and the other is the low-density
growth rate(s) of the invader(s).

\begin{figure}[H]
\includegraphics[width=0.45\textwidth]{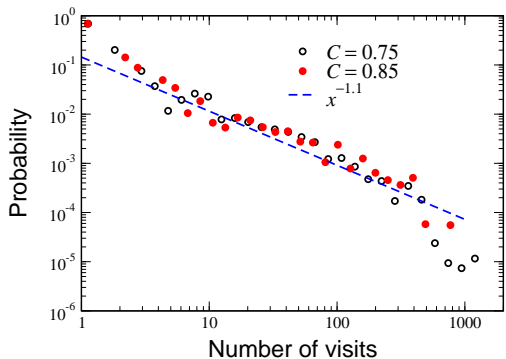}
\caption{Probability distribution for the number of visits to the various invadable metastable states in a run of total duration $\lambda t=2\cdot 10^4$, with snapshots taken each $\lambda t=0.08$ for a total of $2.5 \cdot 10^5$ snapshots.  Data is shown for $C=0.75$ and $0.85$. For $C=0.75$, a total of 598 different states where visited, out of a total of 2045 states. $N=20$, $K=100$, $\lambda=0.01$.  Also shown is is the power-law $P(x) \sim x^{1.1}$, which is a good description of the distribution for all but the most visited states.}
\label{fig8}
\end{figure}

Once successfully invaded, the system leaves the local attractor and wanders
around until it finds another local attractor. One may think about
the local attractor as an abstract network, where each local
attractor is a node and two attractors are connected by a link if the
system may jump directly from one of them to the other. 
This network is seen to have an interesting structure. Figure \ref{fig8}
shows the statistics of number of visits per local attractor, which
is very wide and suggests a (cut-off) power law.  There are a relatively few number of hub states which are visited 
in a significant fraction of the transitions, with a large number of states that are visited relatively infrequently. For example, for $C=0.75$, the most frequently visited state is the invadable state $\{6,\,14\}$, which was visited 1351 times out of $37,543$ transitions. The next most visited states had $1231$, $907$, and $732$ visits, respectively. Actually, for $C=0.75$ there are
three stable uninvadable states.  The stable state $\{1,3,15,18,20\}$ however was only visited twice, and each time lasted only 1 snapshot.  The other two stable states were not visited at all.  Thus, these stable states, for the value of $K=100$ we are studying, are dynamically irrelevant.  We shall return to this point later when we discuss the phase at higher $C$ where various stable states do play a significant role.
 
\begin{figure}[H]
\includegraphics[width=0.45\textwidth]{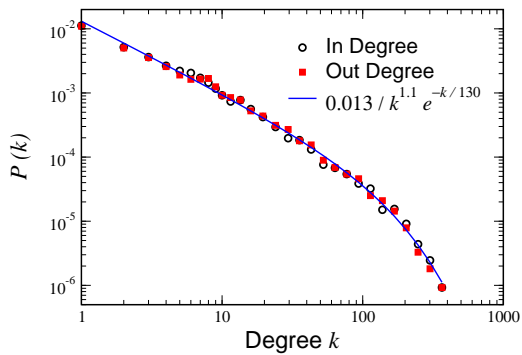}
\caption{Distribution of the in- and out- degrees for the
transitions between the 3729 states encountered in a long run,
$N=40$, $C=0.4$, $K=100$, $\lambda=0.01$.} \label{fig9}
\end{figure}

 The power-law distribution of numbers of visits suggests an
interesting transition network,  with a very heterogeneous structure. Quantifying the
degree distribution of the emerging network we have found that is
quite close to be scale-free.  It shows a power law decay of the
probability of a node to have $k$ links, $P(k)$, with a small
exponent ($ \approx 1.1$), in the case demonstrated in Figure \ref{fig9},
superimposed on a slow exponential cutoff.

\section{Strong competition and Phase IV: the glass-like phase}

We have already mentioned that above $C_2$,  nonstationary attractors can coexist with one or more stable stationary states.  However these 
stationary states had small basins of attraction in addition to containing species with small abundances.  Thus, in the stochastic dynamics they were visited infrequently and had short lifetimes (at least for $K$'s less than $10^4$ or so) and so their dynamical relevance was marginal.  This situation changes qualitatively beyond some larger value of $C$, $C_3$.  Thus, for example, for our standard instance of a $Q=20$, $\sigma=1/2$ $c_{i,j}$ matrix, the stable state 
$\{3,4,6\}$ appears above $C=0.836$ and the stable state $\{2,\,16\}$ appears above $C=0.877$.  The deterministic solution reveals that the smallest abundance in the former state at $C=0.9$ is $0.28K$ and in the latter state, $0.80K$.  For $K=100$, then, they are fairly robust to demographic noise and have correspondingly long lifetimes. These lifetimes are controlled by $K$ and only weakly impacted by decreasing $\lambda$. Due to the small number of species represented in these states, they also are visited relatively frequently.  As $C$ increases further, the depth of the various stable solutions increases nonuniformly, and one state eventually dominates. We operationally define the onset of this fourth phase as the value of $C$ in which the system dwells in a single state for over $20\%$ of the time.  The last phase boundary is illustrated in Fig. \ref{phase}.
%

\section{Encounter at the Hubbell point}

As indicated in Figure \ref{phase}, the transition lines between 
the various phases appear to  meet at the Hubbell point. Clearly, for any
fixed $C<1$ the deterministic model supports full coexistence when
$\sigma \to 0$, implying that the May line must hit the Hubbell point. Similarly, for fixed $C>1$ and vanishingly small
$\sigma$ every solution with one species of abundance $K$ is 
uninvadable, so the line separating the
``chaotic" phase from the glass-like phase also has to reach to
 the Hubbell point. Clearly, it is difficult to implement our operational procedures for determining the phase boundaries in the vicinity of the Hubble point, due to the weak stability of the attractive manifolds in this region, which the noise will smear out.  However, Fig. \ref{phase} indicates the merging of the May line and that separating the partial coexistence and the ``chaotic" phases moves to smaller $\sigma$ for increasing $K$. If the latter boundary indeed extends down to $\sigma=0$ for larger enough $K$, it must also hit the Hubbell point. 

The theory of Hubbell, assuming strict neutrality of all species in
the community, was criticized for this unrealistic assumption. In
particular, it was stressed that any deviation from a strict
neutrality must lead to a fixation of the system by the fittest
species \cite{zhou2008nearly}. As we see here, the situation is more
complicated. $C=1,\  \sigma=0$ is apparently a quadracritical point, with slight
deviations from perfect neutrality yielding different results,
depending on the ratio between $C$ and $\sigma$. When superimposed
on the effect of noise (and, in particular, of demographic
stochasticity that allows for the quasi-absorbing states) the phase
diagram may be very rich.

\begin{figure}[H]
\includegraphics[width=0.45\textwidth]{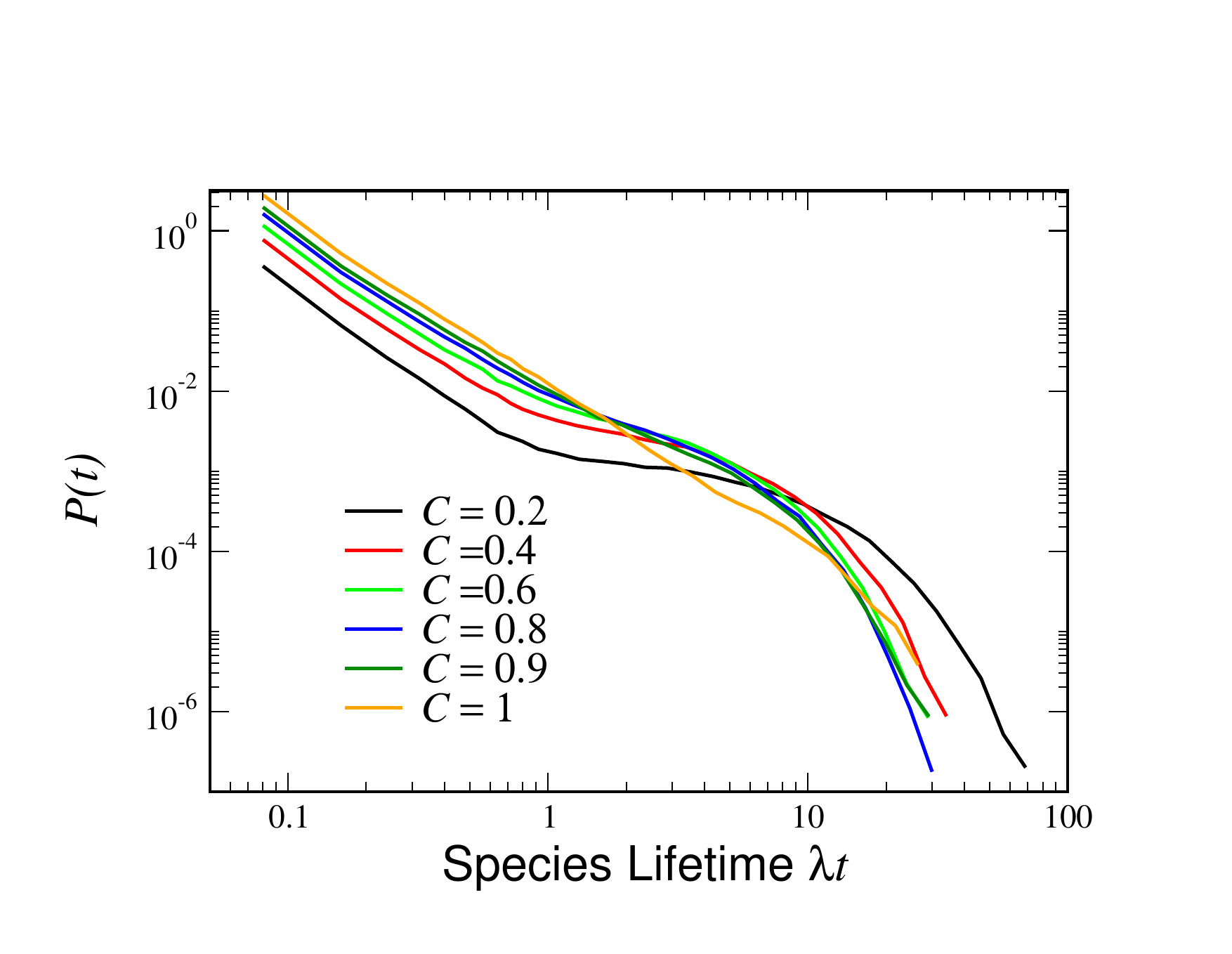}
\caption{Distribution of species lifetimes, for $\sigma=0.0156$,
$K=1000$, $\lambda=0.01$ for various $C$.} \label{fig11}
\end{figure}

One particular example is the distribution of persistence
(colonization to extinction) times of species (as opposed to states, which are
characterized by a given set of extant species). Figure \ref{fig11} shows
this distribution slightly above the Hubbell point, i.e., for $C=1,
\ \sigma = 0.0156$. We see that the distribution of species lifetimes
is quite wide. Indeed, Figure \ref{fig11} suggests a power-law
distribution with an exponent close to 2, which resembles the
findings of \cite{bertuzzo2011spatial}.

\section{Discussion and concluding remarks}

The mainland-island system, considered herein, is one of the main models 
of spatial ecology. It played an important role in the empirical assessment of  the two leading pictures of neutral dynamics, 
the Wilson-MacArthur model (in which all species are equivalent, i.e., admit the same extinction/recolonization rates) and the Hubbell neutral biodiversity model (where all {\em{individuals}} are demographically equivalent). In particular the success of Hubbell in explaining the species abundance distribution in tropical forests, which is the main achievement of the neutral theory of biodiversity, depends entirely on the spatial features of the model, i.e., on the mainland-island structure. The species abundance distribution of the well-mixed model does not fit the empirical data.

In light of this, the rich structure of the mainland-island competitive Lotka-Volterra system that  revealed itself in this study appears to be very interesting. Clearly, the perfect neutrality of Wilson-MacArthur and Hubbell models depends on unrealistic fine tuning of the system parameters, so one would like to figure out what happens when this assumption is relaxed. It turns out that the answer of this question is quite subtle, in particular for the Hubbell point. Slight deviations from the perfectly neutral scenario may lead to absolutely different dynamical behaviors, and the role of noise close to the transition is crucial. 

Our work opens up a few interesting questions about the dynamics of local and global communities. First, one would like to characterize the dynamics of empirical communities as belonging to one of the four qualitative phases considered above. With databases like the North-American Breeding Birds survey, giving the yearly community composition in thousands of locations along about 45 years, this task may be achievable. Once the dynamics of a local community is understood, the overall species turnover rates in a system of local patches connected by migration (a metapopulation) may be investigated both theoretically and by an analysis of field data. One possibility that emerges from our study is that, in such a metacommunity of chaotic or glassy patches, the time to extinction of an extinction-prone species will be so large that it will reach the evolutionary scale (the speciation time) and thus the biodiversity puzzle will be solved. 
           
Two  technical points also merit some discussion. First, although the noise introduced into the model is purely demographic, i.e., it scales with the square root of the population size, the abundance fluctuations are much larger, as clearly seen in Figure \ref{fig all}. The reason is that the demographic noise is superimposed on the nonlinear effects of the deterministic dynamics. This phenomenon is in agreement with many recent studies \cite{Kalyuzhny2014temporal,chisholm2014temporal,Kalyuzhny2014niche}, showing that the noise in empirical systems is clearly stronger than
 demographic. Moreover, at least for large $K$ one should expect that the large abundance semi-resident species are less affected by the noise than the 
 rare species, such that the scaling of abundance fluctuations with $N_i$ will be stronger than the square root of $N_i$ but weaker than $N_i$; this is 
 indeed the case in some empirical systems \cite{Kalyuzhny2014temporal}. 

A second issue, somehow connected to the first, is the effect of ``real" environmental noise, i.e., time dependent fluctuations of  the model parameters. Environmental stochasticity is usually considered as a destabilizing factor, increasing species turnover rate and the amplitude of abundance fluctuations, but it may also stabilize a $c_{i,j}$ independent equal abundance fixed point due to the storage effect \cite{chesson1994multispecies}. We hope to address this issue in subsequent publication. 

Finally,  we believe that the classification  presented here, although only semi-qualitative at present, is very important to the understanding of community dynamics in general. In most cases the data analyzed by researchers  reflect the local  species richness rather than the state of a regional pool, but is interpreted as a fairly honest sample of the global community, assuming, more or less, that the system is either in the heterogeneous coexistence or in the partial coexistence phase. Such an interpretation may be misleading. In particular,  in the chaotic and in the glassy phase sudden drastic  variations in the structure of the community reflect the intrinsic dynamics of the system and, in contrast to a very common interpretation,  are not evidence for exogenous factors that  induce a catastrophic shift. As the concerns about the  impact of anthropogenic changes  rise, it is imperative  to take this possibility into account.

\bibliography{island_ref}

\end{document}